\documentclass[twocolumn,showpacs,preprintnumbers,amsmath,amssymb]{revtex4}
\usepackage{latexsym,epsfig}
\usepackage{amstext}
\usepackage{graphics} 
\usepackage{color}
\usepackage{dcolumn}
\usepackage{bm}

\date{}

\begin{document}

\title{Mirror matter, inverse seesaw neutrino masses and the Higgs mass spectrum}
\author{ M. M. Candido, Y. A. Coutinho, P. C. Malta, \\ J. A. Martins Sim\~oes, A. J. Ramalho}
\email{yara@if.ufrj.br,simoes@if.ufrj.br,ramalho@if.ufrj.br}
\affiliation{ Instituto de F\'isica--Universidade Federal do Rio de Janeiro\\
Av. Athos da Silveira Ramos 149 \\
Rio de Janeiro - RJ, 21941-972, Brazil}

\date{}
\begin{abstract}

In this work we study a mirror model with inverse seesaw neutrino masses in which symmetry breaking
scales are fixed from bounds in the neutrino sector. The Higgs sector of the model has two doublets
and neutral singlets. The mirror model can be tested at the LHC energies in several aspects. Two very
distinctive signatures of the mirror model are a new neutral gauge boson $Z^{\prime}$, with a high invisible branching ratio, 
and a heavy Majorana  neutrino production through the decay $Z^{\prime} \rightarrow N +\bar \nu$.
This result is compared with heavy Majorana production through heavy 
pair production and the consequent same-sign dilepton production.
The other important consequence of the mirror model is the prediction of the Higgs mass. A
particular solution leads to  a Higgs in the same region as in the standard model. There is, however, another
natural solution where the Higgs mass is above $400$ GeV.

\vskip 0.5 cm

\end{abstract}
\maketitle
\eject
\section{Introduction}

\par
Progress in neutrino oscillation experiments gradually confirms that neutrinos are massive and oscillate \cite{PDG}. However, the theoretical understanding of the origin of the mixing pattern and the smallness of neutrino masses has not yet been settled. Many suggestions on possible models for neutrino mixing and masses have been made. For example, the T2K data \cite{T2K} on
$ \sin^2 \, {2 \theta_{13}} > 0$ has motivated models on discrete flavor groups and corrections to the original tri-bimaximal
mixing \cite{ALT}. The MiniBooNE antineutrino data \cite{MIN} has renewed the interest on sterile neutrinos \cite{GIU} and 
extra Higgs doublets can also be a source of new neutrino properties \cite{JAP}.
\par
Neutrino masses and oscillations seem to require new physical scales that are not present in the standard model (SM). There are at 
least three new scales involved: the neutrino mass scale, the lepton number violation scale and the parity breaking scale. All these
scales enter in one of the most appealing extensions of the SM, the left-right symmetric models \cite{JCP}. These 
models  start from the simple gauge structure of $SU(2)_{L}\otimes SU(2)_{R}\otimes U(1)$ and can, hopefully, 
be  tested at the LHC energies. Parity can be broken at the $SU(2)_R$ scale. But it can also be broken  by a  neutral singlet sector, as in the $D$-parity mechanism  developed by Chang, Mohapatra and Parida \cite{CMP}. Small neutrino masses can be generated by the seesaw mechanism.
In this case, lepton number violation is introduced by Majorana terms at very high (GUT) energies. An alternative is the inverse seesaw mechanism \cite{MMV}.
In the original version of the inverse seesaw mechanism, a new left-handed neutrino singlet is introduced. If one imposes lepton 
number conservation in this sector, there are no Majorana mass terms. A new right-handed neutral fermion singlet is also present 
and it is allowed to violate lepton number at a very small scale. This small scale is responsible for the small neutrino mass. 
In this scenario no ultrahigh breaking scale is introduced.

From another point of view, mirror models have recently \cite{MART} been studied  and  it was shown that three additional mirror families are consistent with the standard model if one  additional inert Higgs doublet is included.

This paper is organized as follows: in Section II  we summarize the scalar content of the model. In Section III we present the 
fundamental fermionic representation of the model. In Section IV we discuss the gauge interactions and identify the neutrino 
fields and new $Z^{\prime}$ interactions. In Section V we present the model predictions for the LHC energies. Finally we
summarize the model and its phenomenological consequences in Section VI.

\section{The Higgs scalars}\setcounter{equation}{0}
The fundamental scalar  representation in our mirror model contains the following Higgs scalars: two doublets $\Phi_L$ and $\Phi_R$, which develop the vacuum expectation values $ v_L $ and $v_R$ respectively,
\begin{eqnarray}
\begin{array}{ccc}
& \Phi_L=\left(\begin{array}{cc}
\phi_{_L}^+\\
\\
\phi_{_L}^0\end{array}\right),&
\phi_{_R}=\left(\begin{array}{cc}
\phi_{_R}^+\\
\\
\phi_{_R}^0\end{array}\right),
\end{array}
\end{eqnarray}

\noindent
where
$$\Phi_{_R} \buildrel {\rm D} \over \longleftrightarrow \Phi_{_L}   $$

\noindent with transformation properties under $SU(2)_L\times SU(2)_R \times U(1)_Y$ given by  $(1/2,0,1)_{\phi_L},\; (0,1/2,1)_{\phi_R}$.

The singlet fields of the model are $S_M$, which develops a v.e.v. at a very small scale and is coupled with Majorana mass terms,
and $M_{N_L}$, $M_{N_R}$, which must couple to lepton number conserving terms (Dirac) at a TeV scale.

For the lepton number violating singlet we impose the symmetry,
$$S_M \buildrel {\rm D} \over \longleftrightarrow S_M $$
and for the lepton number conserving singlets,
$$M_{N_L} \buildrel {\rm D} \over \longleftrightarrow -M_{N_R}. $$
\par
These scalar fields will develop vacuum expectation values according to
$${\phi_L,\phi_R,S_M,M_{N_L},M_{N_R}} \buildrel {\rm v.e.v.} \over \longleftrightarrow {v_L,v_R,s,v_{M_L},v_{M_R}}. $$
\par
The motivation behind these symmetries is to generate a simple spectrum for the neutrino sector ( see section III ). The $\phi_L$ field will be broken at the same scale of the SM Higgs field $v_L = v_{\, \text{Fermi}}$. The new $v_R$ scale can be searched 
for at the LHC energies  in the $ 1-10 $ TeV range. The bound from neutrinoless double beta decay will imply ( see  section IV ) that $v_{M_L} > 1$ TeV and $v_{M_R} > 10^5 $ TeV. The $S_M$ singlet field will break lepton number at a  small scale $s \simeq 1$ eV and will give small neutrino masses.
\par
 The most general scalar potential invariant under the preceding symmetries  has more than twenty new parameters. We can obtain constrained equations and stability conditions  from a simpler form, still consistent with the 
stated symmetries:
\begin{widetext}
\begin{eqnarray}
V &=& -\mu_1^2 S_M^2 -\mu_2^2( M_{N_L}^2 +M_{N_R}^2) -\mu_3^2 (\phi_L^2 + \phi_R^2)+\mu_{4}^2M_{N_L}M_{N_R} + 
\mu_{5} ( M_{N_L} -M_{N_R})(\phi_L^2 - \phi_R^2) \nonumber\\
&+& \lambda_1S_M^4 +\lambda_2( M_{N_L}^4 +M_{N_R}^4)+ \lambda_3 M_{N_L}^2 M_{N_R}^2 + \lambda_4(\phi_L^4 + \phi_R^4) + \lambda_5\phi_L^2 \phi_R^2 .
\end{eqnarray} 
\end{widetext}

The $\phi_{L,R}$ doublets will give masses to the  gauge bosons of $ SU(2)_{L,R}$ respectively. There will remain five neutral scalar Higgs fields in the model. It is straightforward, although lengthy, to find the constraint equations and Hessian matrix that guarantee the minimum conditions. They are explicitly given in the Appendix.

The approximate eigenvalues of the squared-mass matrix are given by
\begin{eqnarray}
&& M_1^2 = 8 \lambda_1 s^2 \nonumber\\
&& M_2^2 \simeq 4\left[ \lambda_4 (v_R^2+v_L^2) - \sqrt{ \lambda_4^2 (v_R^2-v_L^2)^2 + \lambda_5 ^2 \, v_L^2 v_R^2 }\right] \nonumber\\
&& M_3^2\simeq 4\left[ \lambda_4 (v_R^2+v_L^2) + \sqrt{ \lambda_4^2 (v_R^2-v_L^2)^2 + \lambda_5^2 \, v_L^2 v_R^2 }\right] \nonumber\\
&& M_4^2\simeq \left[ \lambda_3 + 2 \lambda_2 - \vert \lambda_3-6 \lambda_2 \vert \right] v_{M_R}^2\nonumber\\
&& M_5^2\simeq \left[ \lambda_3 + 2 \lambda_2 + \vert \lambda_3-6 \lambda_2 \vert \right] v_{M_R}^2
\end{eqnarray}

The most prominent feature of these expressions is the prediction for the squared-mass value that corresponds to the standard model 
Higgs. This result for the Higgs mass shows that mirror models have a  clear difference with the standard model Higgs.
The recent LHC experimental searches for the standard model Higgs  have detected no positive signal. There are increasing data 
constraining some regions for the Higgs mass value \cite{FERM}. According to the recent data from ATLAS \cite{ATLAS} and CMS \cite{CMS} 
collaborations, there still remains a first open window for the Higgs mass in the $116-145$ GeV region. This limits the free parameters 
of our invariant potential to the following region in parameter space:

\begin{eqnarray}
&& 0 < \lambda_1 < 1      \nonumber\\
&& 0 < \lambda_2 < 1/2     \nonumber\\
&& 0 < \lambda_3 < 1       \nonumber\\
&& 0.05 < \lambda_4 < 0.56 \nonumber\\
&& -1 < \lambda_5 < 1
\end{eqnarray}

The ATLAS and CMS collaborations also exclude with $95\%$ C.L. the existence of a Higgs over most of the mass region from $145$ to $466$
GeV. A second open mass window $ M_{\text Higgs} > 466 $ GeV  would imply the same values for $\lambda_i$, except for $ \lambda_4 > 0.8 $.

\section{Neutrinos in the mirror model}
The fundamental fermion representation for the first lepton family and its transformation  under the discrete parity 
symmetry ($D$ parity) in the mirror model is given by
\begin{equation}
\left(
\begin{array}{c}
\nu \\
e \\
\end{array}\right)_L,\
\nu_{_R},\
e_{_R} \quad \buildrel {\rm D} \over \longleftrightarrow \quad
\left(
\begin{array}{c}
N \\
E \\
\end{array}\right)_R,\
N_{_L},\
E_{_L},
\end{equation}\

\noindent
where the doublets transform under $SU(2)_L\times SU(2)_R \times U(1)_Y$ as $(1/2,0,-1)_{L},\; (0,1/2,-1)_{R}$.

In order to discuss the  mass for  the neutral fermion fields we start by considering the following Majorana fields coming from the 
fundamental mirror representation:
\begin{eqnarray}
\Psi_{\nu} \equiv \nu_{_L} + \nu_{_L}^c  \qquad  \Psi_{_N} \equiv N_{_L} + N_{_L}^c \nonumber\\
\omega_{\nu} \equiv \nu_{_R} + \nu_{_R}^c  \qquad  \omega_{_N} \equiv N_{_R} + N_{_R}^c.
\end{eqnarray}
The doublets transform as  $(1/2,0,-1)_{L},\; (0,1/2,-1)_{R}$.
Let us discuss the mass Lagrangian by showing explicitly the physical content of each term.
The  mirror mass Lagrangian coupled with the Higgs doublets is given by

\begin{eqnarray}
\mathcal{L}_{M}^{(\text{mirror})} &=& \frac{1}{2} v_L [ { \bar\nu_{_L} \nu_{_R} + \bar\nu_{_R} \nu_{_L} + \bar{\nu^c_{_L}} \nu^c_{_R} + 
\bar{\nu^c_{_R}} \nu^c_{_L}} ] \nonumber\\
&+& \frac{1}{2} v_R [\bar N_{_L} N_{_R} + \bar N_{_R} N_{_L} + \bar {N^c_{_L}} N^c_{_R} + 
\bar{N^c_{_R}} N^c_{_L} ]. \nonumber\\
\end{eqnarray} 

In this expression we have no Majorana mass terms that violate lepton number. In the Majorana field basis we have
\begin{eqnarray}
\mathcal{L}_{M}^{(\text{mirror})} &=& \frac{1}{2} v_{_L}[ \bar\Psi_{\nu} \,\omega_{\nu} + \bar\omega_{\nu} \,\Psi_{\nu}] 
+
\frac{1}{2}v_R [ \bar\Psi_{_N} \,\omega_{_N} + \bar \omega_{_N} \,\Psi_{_N}].\nonumber\\
\end{eqnarray} 

As required by the inverse seesaw mechanism we must introduce a new neutral fermionic singlet (called "$P$"). As we are considering a
parity conserving model  both left and right handed components of this field must be present. We have a new Lagrangian mass term given by:

\begin{eqnarray}
\mathcal {L}_{M}^{P} &=& \frac{s}{2} [ \bar P_{_L} P^c_{_L} +  \bar P^c_{_L} P_{_L}  + \bar P_{_R} P^c_{_R} +  \bar P^c_{_R} P_{_R} ] \nonumber\\
&+& v_{M_L}  [\bar\nu_{_R}P_{_L} + \bar P_{_L} \nu_{_R} ] 
- v_{ M_R}  [\bar N_{_L} P_{_R} + \bar P_{_R} N_{_L}] \nonumber\\
&+& \frac{s}{2}  [ \bar P_{_L} N^c_{_L} + \bar N^c_{_L} P_{_L}  + \bar P^c_{_L} N_{_L} + \bar N_{_L} P^c_{_L} ] \nonumber\\
&+& \frac{s}{2}  [ \bar P_{_R} \nu^c_{_R} + \bar \nu^c_{_R} P_{_R} + \bar P^c_{_R} \nu_{_R} + \bar \nu_{_R} P^c_{_R} ].
\end{eqnarray} 

We have now new Majorana fields,
\begin{eqnarray}
\epsilon \equiv  P_{_L} + P^c_{_L} \qquad  \sigma = P_{_R} +  P_{_R}^c,
\end{eqnarray}
and these terms give a new contribution to the mass Lagrangian ,

\begin{eqnarray}
\mathcal {L}_{M}^{P} &=& \frac{s}{2} [ \bar\epsilon \,\, \sigma +  \bar \sigma \,\, \epsilon 
+ \bar \epsilon \,\, \epsilon  +  \bar \sigma \,\, \sigma ] \nonumber\\
&+& v_{M_L}  [\bar \omega_{\nu} \,\, \epsilon + \bar \epsilon \,\, \omega_{\nu} ] 
- v_{M_R}  [ \bar \Psi_{_N} \,\, \sigma + \bar \sigma \,\, \Psi_{_N}  ] \nonumber\\
&+& \frac{s}{2}  [ \bar \epsilon \,\, \Psi_{_N} + \bar \Psi_{_N} \,\, \epsilon  +
 \bar \sigma \, \omega_{\nu} + \bar \omega_{\nu} \,\, \sigma ].
\end{eqnarray}

Returning now to the Majorana basis, the full mass Lagrangian can be written as 
\begin{widetext}
\begin{eqnarray} 
\mathcal{L}_{\text mass}
&=& \left( \bar{\Psi_{\nu}} \,\, \bar \omega_{\nu} \,\,  \bar \epsilon \,\,  \bar \omega_{_N} \,\, \bar \Psi_{_N} \,\, 
\bar \sigma \right)
\left(\begin{array}{cccccc}
0 &  v_{_L}/2  & 0 & 0  & 0 & 0 \\
\\
v_{_L}/2  & 0 & v_{M_L} & 0 & 0 &  s/2 \\
\\
0 & v_{ M_L}  & s/2  & 0 &  s/2 &  0   \\
\\
0  & 0 & 0  & 0  & v_{_R}/2  & 0 \\
\\
0 & 0 & s/2 & v_{_R}/2 & 0 & -v_{M_R} \\
\\
0 & s/2 & 0 & 0 &  -v_{M_R} & s/2
\end{array} \right)
\left(\begin{array}{c}
\Psi_{\nu}\\
\\
\omega_{\nu}\\
\\
\epsilon\\
\\
\omega_{_N}\\
\\
\Psi_{_N} \\
\\
\sigma
\end{array} \right).\nonumber
\end{eqnarray}
\end{widetext}
This last matrix has two blocks  in the inverse seesaw form,
\begin{eqnarray} 
M
&=& 
\left(\begin{array}{ccc}
0 & v  & 0 \\
v & 0 & M \\
0 & M & s 
\end{array} \right).\nonumber
\end{eqnarray}
As $ s $ will be responsible for the very small neutrino masses, it must have a very small value. Then the general mass 
matrix, to first order, has two independent inverse seesaw blocks.

The diagonalization  of the mass matrix, to first order, allows to calculate the mass eigenvalues and eigenvectors. 
Introducing the notation
\begin{eqnarray}
&& R_{_L} \equiv \sqrt{v_{_L}^2+v_{M_L}^2}, \quad
R_{_R} \equiv \sqrt{v_{_R}^2+v_{M_R}^2}, \nonumber\\
&&
s_{_L} = \frac{v_{_L}}{R_{_L}}, \quad c_{_L} \equiv  \frac{v_{M_L}}{R_{_L}}, \nonumber\\
&&
s_{_R} = \frac{v_{_R}}{R_{_R}}, \quad c_{_R} \equiv  \frac{v_{M_R}}{R_{_R} },
\end{eqnarray}

\begin{widetext}
\begin{eqnarray}
m_1 \simeq R_{_L}  \quad \qquad m_2 \simeq - R_{_L} \quad \qquad m_3 \simeq s^2_{_L}\,\, s
\end{eqnarray}
\begin{equation}
\Psi_1= \left(
\begin{array}{c}
\frac{{\sqrt 2}}{2} {s_{_L}}\\
0 \\
\frac{\sqrt{2}}{2}\\
0 \\
-\frac{{\sqrt 2}}{2} {s_{_L}} \\
0
\end{array}\right)\nonumber\\
\qquad
\Psi_2= \left(
\begin{array}{c}
\frac{{\sqrt 2}}{2} {s_{_L}}\\
0 \\
-\frac{\sqrt{2}}{2}\\
0 \\
-\frac{{\sqrt 2}}{2} {s_{_L}} \\
0
\end{array}\right)
\qquad
\Psi_3= \left(
\begin{array}{c}
c_{_L} \\
0 \\
-s_{_L} c_{_L} \frac{s}{R_{_L}}\\
0 \\
-s_{_L} \\
0
\end{array}\right)
\end{equation}

\begin{eqnarray}
m_4 \simeq R_{_R}  \qquad \qquad m_5 \simeq - R_{_R} \qquad \qquad m_6 \simeq s^2_{_R}\,\, s
\end{eqnarray}

\begin{equation}
\Psi_4= \left(
\begin{array}{c}
0 \\
\frac{{\sqrt 2}}{2} {s_{_R}}\\
0\\
\frac{\sqrt{2}}{2}\\
0 \\
-\frac{{\sqrt 2}}{2} {c_{_R}} \\
0
\end{array}\right)\nonumber\\
\qquad
\Psi_5= \left(
\begin{array}{c}
0 \\
\frac{{\sqrt 2}}{2} {s_{_R}}\\
0\\
-\frac{\sqrt{2}}{2}\\
0 \\
-\frac{{\sqrt 2}}{2} {c_{_R}} 
\end{array}\right)
\qquad
\Psi_6= \left(
\begin{array}{c}
0\\
-c_{_R} \\
0 \\
-s_{_R} c_{_R} \frac{s}{R_{_R}}\\
0 \\
-s_{_R} \\
0
\end{array}\right)
\end{equation}

The diagonalization matrix can be written as:
\begin{eqnarray} 
U
&=& 
\left(\begin{array}{cccccc}
\frac{\sqrt{2}}{2} s_{_L} & \frac{\sqrt{2}}{2} s_{_L} &  - c_{_L} & 0  & 0 & 0 \\
0 & 0  & 0 & \frac{\sqrt{2}}{2} s_{_R}  & \frac{\sqrt{2}}{2} s_{_R} & -c_{_R} \\
\frac{\sqrt{2}}{2}  & -\frac{\sqrt{2}}{2} & - s_{_L}c_{_L} \frac{s}{R_{_L}} & 0 & 0 & 0  \\
0  & 0 & 0  & \frac{\sqrt{2}}{2}  & -\frac{\sqrt{2}}{2}  &  s_{_R}c_{_R} \frac{s}{R_{_R}} \\
- \frac{\sqrt{2}}{2}c_{_L}  & - \frac{\sqrt{2}}{2}c_{_L} &  s_{_L} & 0  & 0 & 0 \\
0 & 0 & 0  & - \frac{\sqrt{2}}{2}c_{_R}  & - \frac{\sqrt{2}}{2}c_{_R}  & -s_{_R}
\end{array} \right).\nonumber
\end{eqnarray}
\end{widetext}

\section{The  gauge interactions}
In order to proceed  to the neutrino identification we must look at the neutral current interactions.
The general gauge structure of our model was developed in ref. \cite{DEALM}. From  equation (18), the neutral gauge bosons $Z$ 
and $Z^{\prime}$ interact only with  $\nu_L$ and $N_R$. All other neutrino states have no gauge interactions as they are neutral singlets.

Neglecting $\omega^2$ terms, in the Majorana basis this is given by equation (19) of ref. \cite{DEALM},
\begin{widetext}
\begin{eqnarray}
\mathcal {L}_{NC} &=& \frac{-g_{_L}}{2 \cos\theta_{W}}[ \bar \Psi_{\nu} \frac{\gamma^{\mu}(1- \gamma^5)}{2} 
\Psi_{\nu} ]Z_{\mu}
-  \frac{g_{_L}}{2} \tan\theta_W \tan\beta [ \bar \Psi_{\nu} \frac{ \gamma^{\mu}(1- \gamma^5)}{2} 
\Psi_{\nu} + \frac{1}{\sin^2\beta}  \bar \omega_{_N}  \frac{\gamma^{\mu}(1+ \gamma^5)}{2} \omega_{_N}
] Z^{\prime} _{\mu}. 
\end{eqnarray} 
\end{widetext}

As the $\Psi_{\nu}$ field is given by
$$ \Psi_{\nu} = \frac{\sqrt 2}{2}  [ s_{_L} \Psi_1 + \Psi_3 - c_{_L} \Psi_5 ],$$
\noindent
the relevant combination for the $Z$ interaction comes from
\begin{eqnarray*}
\bar \Psi_{\nu}  \Psi_{\nu}  &\rightarrow &  \frac{1}{2} [ s^2_{_L}  \bar \Psi_1 \Psi_1 +
 \bar \Psi_3 \Psi_3 + c^2_{_L} \bar \Psi_5 \Psi_5  \nonumber\\
&+& s_{_L} \bar \Psi_1 \Psi_3 - s_{_L} c_{_L} \bar \Psi_1 \Psi_5 -
c_{_L}\bar \Psi_3 \Psi_5 ].
\end{eqnarray*}

Hence, the $Z$ full coupling is given by the light $\Psi_3$ state. This state is to be identified with the SM neutrino.

There is no (light-light) $\Psi_3 -\Psi_6$ mixing and the $Z$ width is the same as in the SM. As $\Psi_5$ is the heaviest
state, we will have the leading terms:
$$\bar \Psi_{\nu} \Psi_{\nu} \simeq \frac{1}{2} [ \bar \Psi_3 \Psi_3 + s_{_L} \bar \Psi_1 \Psi_3 ].$$

The $Z^{\prime}$ interaction involves the $\omega_{_N}$ state as
$$\omega_{_N} =  \frac{\sqrt 2}{2} [ \Psi_1 s_{_L} - \Psi_3 - c_{_L} \Psi_5 ],$$
\noindent and we have

\begin{eqnarray*}
\bar \omega_{_N} \omega_{_N} &\rightarrow & \frac{1}{2} [  s^2_{_L} \bar \Psi_1 \Psi_1 + \bar \Psi_3 \Psi_3 + c^2_{_L}\bar \Psi_5 \Psi_5 \nonumber\\
&-& s_{_L} \bar \Psi_1 \Psi_3 - s_{_L}c_{_L}\bar \Psi_1 \Psi_5 + c_{_L} \bar \Psi_3 \Psi_5 ].
\end{eqnarray*}

The leading terms are
$$\bar \omega_{_N} \omega_{_N} \simeq \frac{1}{2} [ \bar \Psi_3 \Psi_3 -s_{_L} \bar \Psi_1 \Psi_3 ].$$

So the new  $Z^{\prime}$ will decay in the light state  $Z^{\prime} \longrightarrow \bar \nu_3 \nu_3$ but with a 
coupling much larger than that of SM case. The $Z^{\prime} \bar \nu_3 \nu_3$ vertex is given by

$$Z^{\prime}\bar \nu_3 \nu_3  \simeq  \frac{g_L}{2} \tan\theta_W \tan\beta (1 + \frac{1}{\sin^2 \beta}).$$

\noindent 
We  also  have the interaction vertex 
$$Z^{\prime} \bar \nu_1 \nu_3  \simeq  \frac{g_L}{2} \tan\theta_W \tan\beta (1 - \frac{1}{\sin^2 \beta}), $$
and this term can also be quite large.

The charged current interaction is given by
\begin{eqnarray}
{\cal L}_{e\, N\, W} &=& \frac{g}{2 \sqrt 2} \bar e_{_L} \gamma^{\mu} \nu_{_L} W_{\mu} \nonumber\\
&=& \frac{g}{2 \sqrt 2} \bar e \gamma^{\mu} [s_{_L} \Psi_1  + \Psi_3 + c_{_L} \Psi_5 ] W_{\mu},
\end{eqnarray}

where $\Psi_3$ is the SM neutrino state.

From neutrinoless double $\beta$ decay ($0\nu\beta\beta$) we have the experimental bound \cite{RODE}
$$\frac{\sin^2\theta_{e\, N\, W}}{M_{_N}} < 5 \times 10^{-8} \, {\rm GeV}^{-1}.$$

\par
For the first heavy neutrino ($\Psi_1$) we obtain the bound 
$$\frac{v_L^2}{v^2_L+v_{M_L}^2} \frac{1}{\sqrt{v^2_L +v_{ M_L}^2}}\simeq \frac{v^2_L}{v_{M_L}^3} < 5 \times 10^{-8}\, {\rm GeV}^{-1},$$

\noindent
which implies $v_{M_L} > 1 - 10$ TeV. This uncertainty comes from the absorption of coupling constants in the definition of
our $ v_i$. If we let the corresponding couplings vary in the range $ g_i \simeq 0.1 - 1 $, then the preceding result follows.

For the second heavy neutrino ($\Psi_5$) we have
$$\frac{v_{M_L}^2}{v^2_L+v_{M_L}^2} \frac{1}{\sqrt{v^2_R + v^2_{M_{R}}}}\simeq \frac{1}{v_{M_{R}}} < 5 \times 10^{-8} \, {\rm GeV}^{-1},$$
so that $ v_{M_R} \gtrsim 10^5 $  TeV.

It is a remarkable result that from neutrino bounds we have recovered  the Peccei-Quinn scale related to the strong CP problem \cite{CAR}.
With the identification  $\Psi_3 \longrightarrow \Psi_{\nu_e}$ and $\Psi_1 \longrightarrow \Psi_N$, the leading new $Z^\prime$ interaction with neutrinos is

\begin{eqnarray}
\mathcal {L}_{NC} &=& -\frac{g_L \tan\theta_{\omega} \tan\beta}{4} \{ \bar\Psi_{\nu_e} \gamma^{\mu} [g_A -g_V \gamma^5] \,
\Psi_{\nu_e}  \nonumber\\
&+& s_L \bar\Psi_{\nu_e} \gamma^{\mu} [g_A -g_V \gamma^5] \Psi_N + h.c.\} \,Z^{\prime}_{\mu}
\end{eqnarray} 
\noindent
where 
$$g_V = 1 - \frac{1}{\sin^2\beta}  \qquad {\rm and} \qquad  g_A = 1 + \frac{1}{\sin^2\beta}.$$
\smallskip

From the preceding relations the $Z^{\prime} {\bar N} N $ vertex is suppressed by an  $ {s_L}^2$ factor.

\section {Results}
In this section we present the main phenomenological consequences of our model for the LHC. Although many extended models
predict a new  $Z^{\prime}$, it is a very distinctive property of mirror models that the invisible $Z^{\prime}$ channel 
will be very high. In Table I we show the branching ratios for the $M_{Z^{\prime}} = 1.5 $ TeV with $\Gamma_{Z^\prime} \simeq 25 $ GeV 
considering  $v_{M_L} = 1 - 10$ TeV. The heavy neutrino channels are strongly dependent on the choice of $v_{M_L}$.

The clearest signal for a new $Z^{\prime}$ will be the leptonic channel $p + p  \rightarrow  l^{+} + l^{-}+ X$. The recent LHC searches for
this process have not detected any evidence of a new $Z^{\prime}$ boson. For instance, the ATLAS Collaboration \cite{ATL} with a luminosity 
around $1$ fb$^{-1}$ sets a lower bound $M_{Z^{\prime}} > 1.83$ TeV on the mass of a new sequential heavy $Z^{\prime}$. Using the package 
CompHep \cite{COMP} with CTEQL1 parton distribution functions, we have estimated the corresponding bound on the mass of the 
mirror $Z^{\prime}$ boson. Applying a set of cuts on the final leptons, namely, $\vert\eta\vert \leq 2.5$, 
$ M_{Z^{\prime}} - 5 \, \Gamma_{Z^{\prime}} < M_{l^- l^+} < M_{Z^{\prime}} + 5 \, \Gamma_{Z^{\prime}}$,  and an energy cut of  
$ E_T > 25$ GeV,  we display in Figure 1a the total cross section and number of events
for  $\sqrt s = 7$ TeV and an integrated luminosity of $10$ fb$^{-1}$. The negative 
result of the ATLAS search leads, therefore, to a bound $M_{Z^{\prime}} > 1.5$ TeV on the $Z^{\prime}$ mass in our model. The forthcoming 
luminosity of $10$ fb$^{-1}$ will allow the search for this $Z^{\prime}$ to be extended up to $M_{Z^{\prime}} = 2.0$ TeV.
In Figure 1b we show our results for a center of mass energy of $14$ TeV and  an integrated luminosity of $ 100$ fb$^{-1}$. In 
this case we can estimate an upper bound on $M_{Z^{\prime}}$ around $4.0$ TeV.

\begin{figure} 
\begin{center}
\includegraphics[width=.45\textwidth]{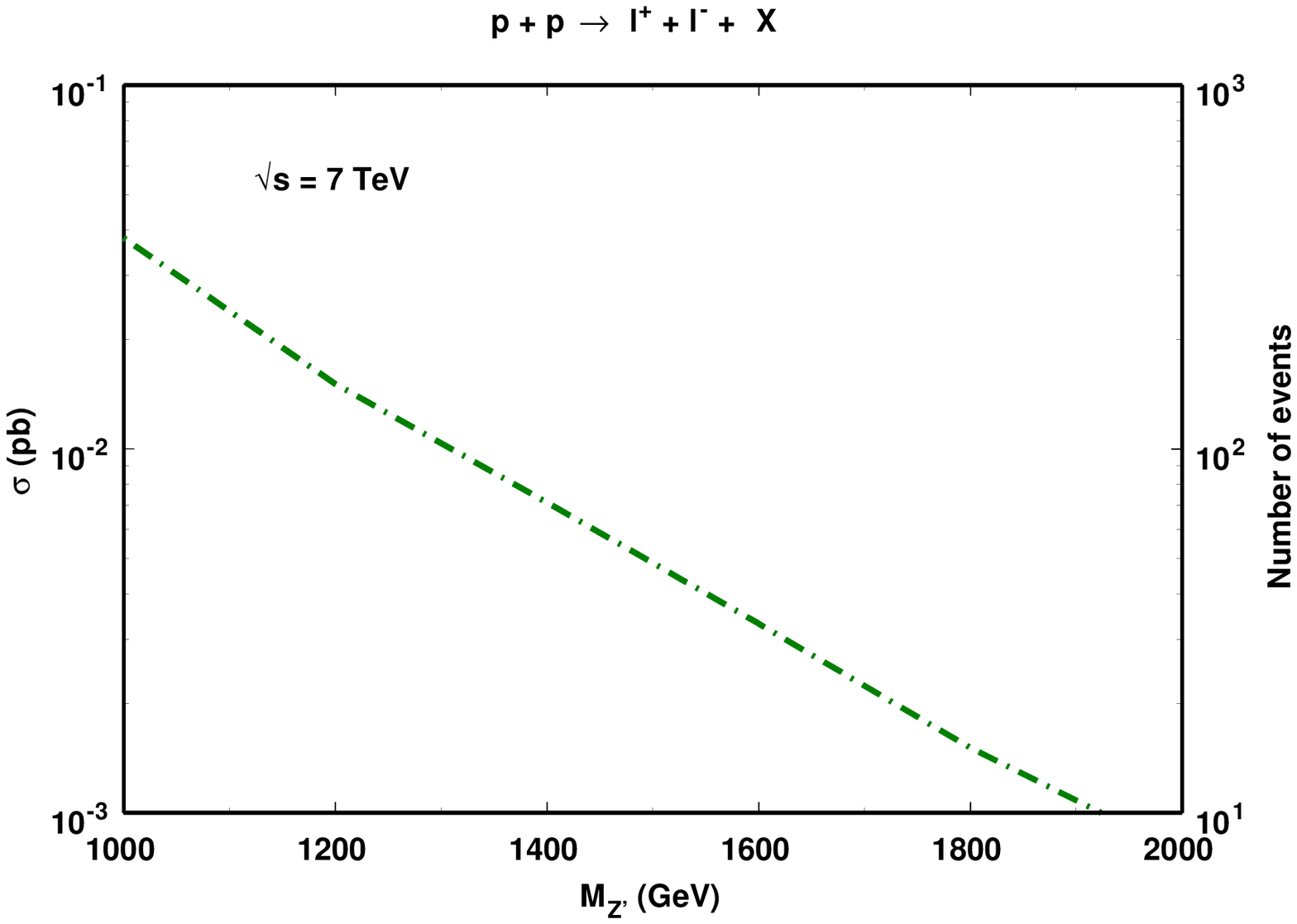}
\includegraphics[width=.45\textwidth]{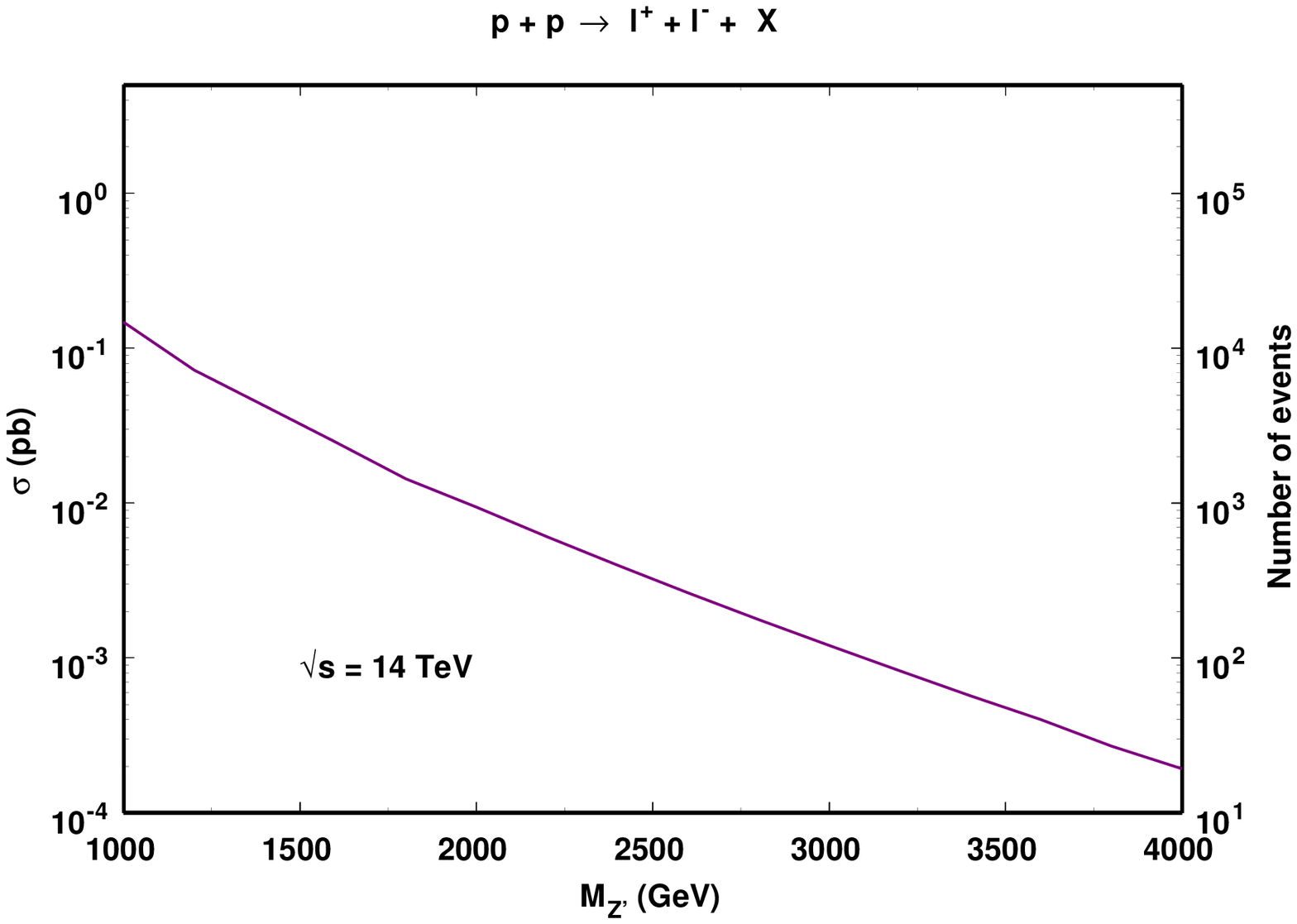}
\caption{Total cross section and number of events versus $M_{Z^{\prime}}$ in the process $p + p  \rightarrow  l^{+}+ l^{-}+ X$ 
at  $\sqrt s= 7$ TeV considering ${\cal L}= 10$ fb$^{-1}$ (up) and at $\sqrt s= 14$ TeV considering ${\cal L}= 100$ fb$^{-1}$ (down).}
\end{center} 
\end{figure}

Another important test of the model is the prediction of heavy Majorana neutrinos with masses up to the TeV region \cite{CHART}. From Table 1 we 
see that the dominant heavy Majorana neutrino production in our model is through $Z^{\prime} \rightarrow N +\bar \nu$. This result is to be 
compared with a very similar model \cite{DEALM} where the neutrino masses comes from a double seesaw mechanism. In this last case, the dominant heavy Majorana production is through heavy pair production and the consequent same-sign dilepton production. In Figure 2 we show the total cross section for the process  $p + p  \rightarrow N +\bar \nu + X$ for the planned LHC energies and luminosities. The final state will be seen as
$p + p \rightarrow  {\rm invisible} +  l^{\pm}+ W^{\mp}+ X$, with the invariant $l^{\pm}+ W^{\mp}$ mass peaked at the 
heavy neutrino mass.
For heavy Majorana neutrinos with masses near $100$ GeV the dominant production 
mechanism is through the SM $W$ and $Z$ interactions. However this mechanism is kinematically restricted to masses below 
$200$ GeV \cite{ALM}. For higher masses the dominant mechanism is {\it via}  $Z^{\prime}$ exchange. From Figure 2  we can estimate 
the heavy neutrino mass dependence at the energy of $\sqrt s = 7$ TeV produced {\it via}  $Z^{\prime}$ with mass equal to 
$ 2.0 $ TeV. The scenario of $\sqrt s = 14$ TeV allows us to estimate the $M_N$ behavior from $500$ GeV to 
$2$ TeV, with $Z^{\prime}$ masses varying from $1.5$ TeV to $3.0$ TeV.

\par
\begin{figure} 
\begin{center}
\includegraphics[width=.45\textwidth]{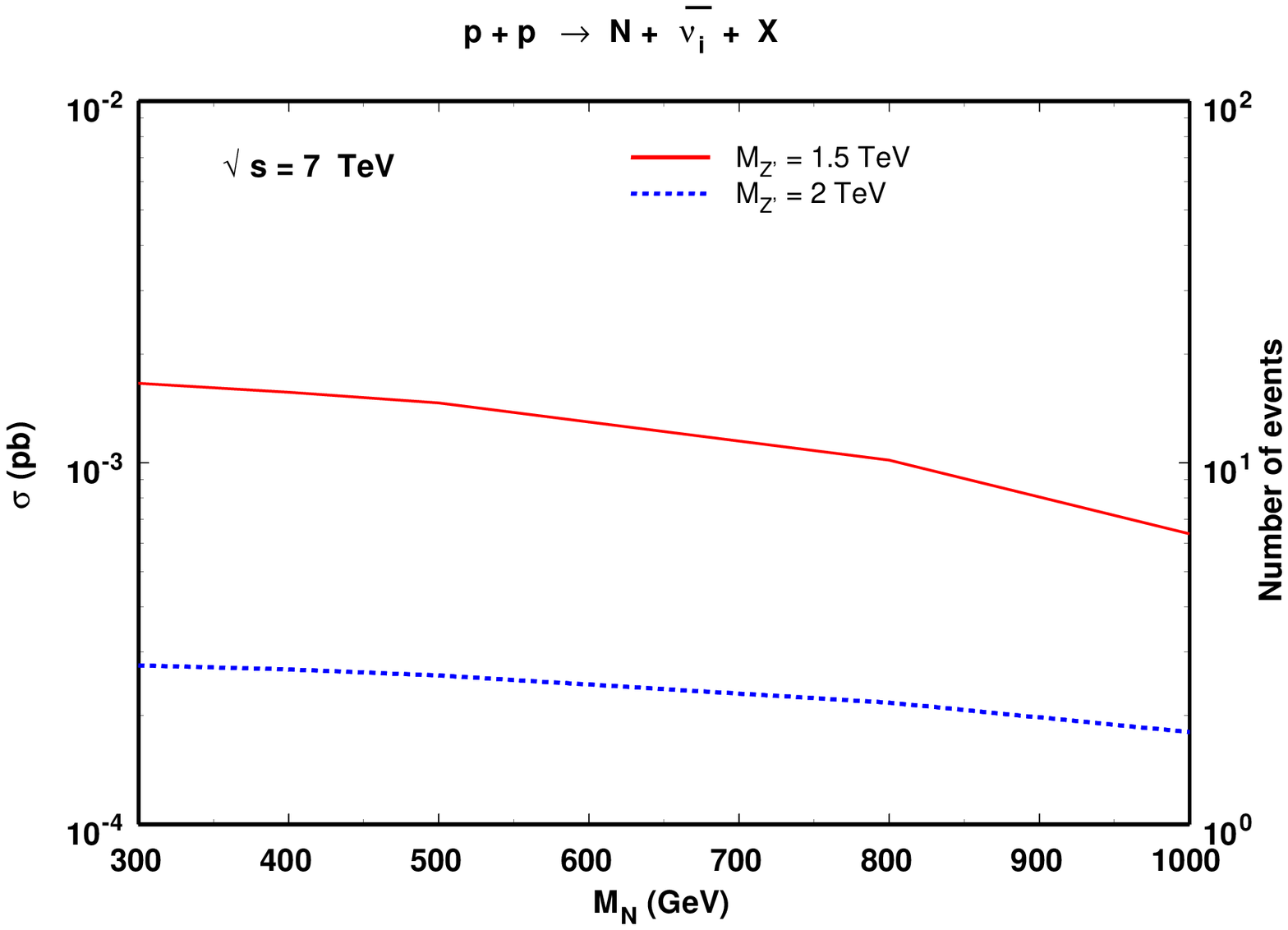}
\includegraphics[width=.45\textwidth]{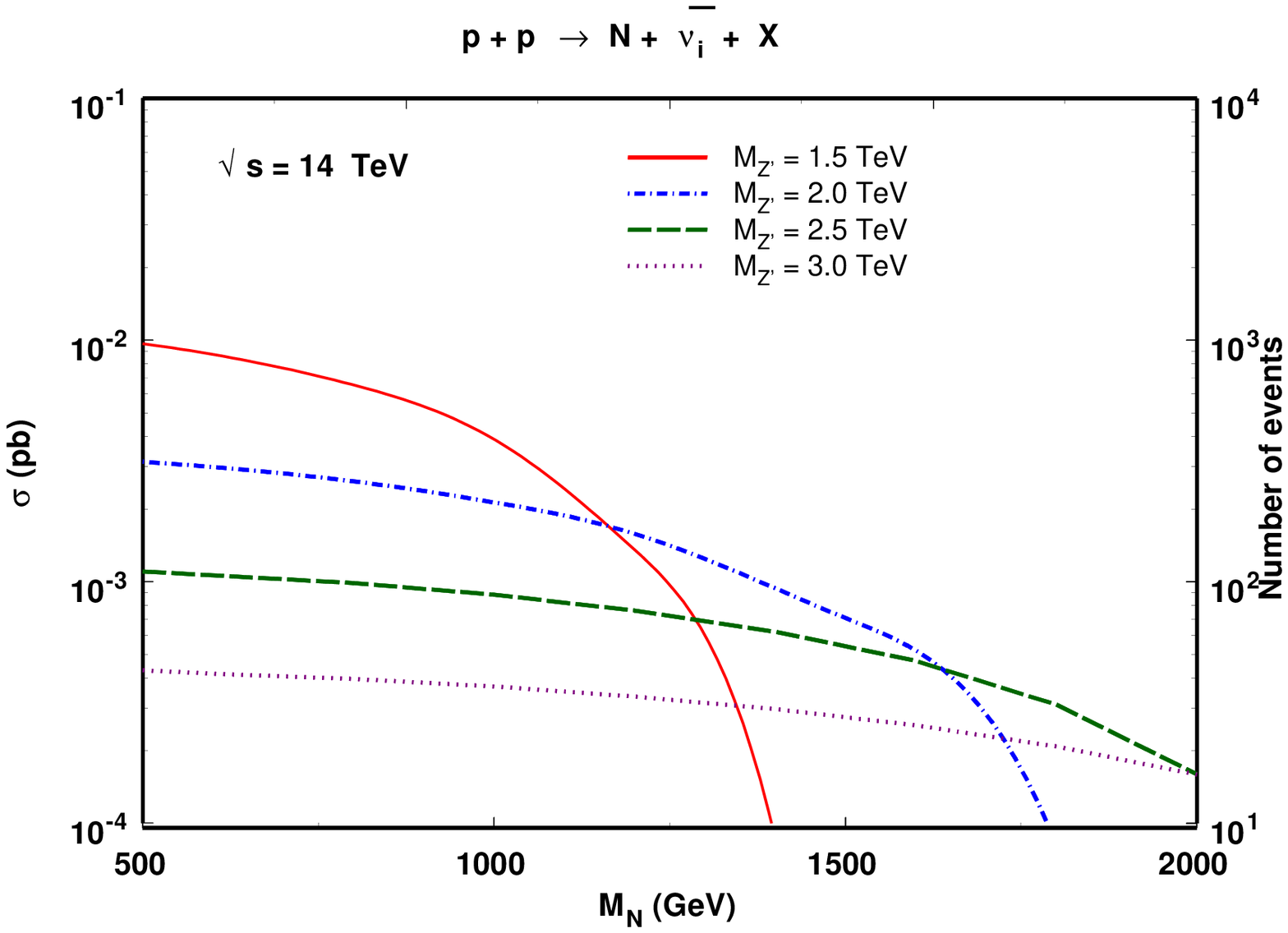}
\caption{Total cross section and number of events versus $M_N$ in the process $p + p  \rightarrow   N + \bar \nu_i+ X$ 
at $\sqrt s= 7 $ TeV considering ${\cal L}= 10$ fb$^{-1}$ (up) and  at $\sqrt s= 14$ TeV considering ${\cal L}= 100$ fb$^{-1}$ (down).}
\end{center} 
\end{figure}

{\section {Conclusions}}
 In this paper we have presented a mirror model that restores parity at high energies. Neutrino masses are generated by 
the inverse seesaw mechanism. Besides new mirror fermions, the model also predicts new gauge vector bosons. Our choice of a scalar sector with 
Higgs doublets and singlets and no Higgs bidoublets means that the new charged vector bosons will not be coupled with
ordinary matter in the mirror model at tree level.  This points to a significant difference between  the present model and other left-right models 
with new $\nu_R$ neutrinos in new $SU(2)_R$ doublets . But mixing in the neutral vector boson sector is present 
and the first important phenomenological consequence of the model is a new neutral current. As the new v.e.v. $v_R$ is not known,
we cannot determine exactly the new $Z^{\prime}$ mass. But the LHC can test the hypothesis that $v_R$ is of the order of a few TeV. 
The new $Z^{\prime}$ mixing with the other neutral gauge bosons can be calculated \cite{PON} and we can determine both the main 
decay channels and production rates for this new $Z^{\prime}$. 

The heavy Majorana neutrino production  can be used as a test for the basic neutrino mass generation mechanism. 
In the double seesaw mechanism we have the dominant channel $ Z^{\prime} \rightarrow N +  N$ and the consequent same-sign 
dilepton production, whereas for the inverse seesaw mechanism the dominant channel is $ Z^{\prime} \rightarrow N +\bar \nu$.

The other important prediction of our mirror model comes from the fact the we have fixed the symmetry breaking scales only from the neutrino 
sector; therefore, the Higgs spectrum can be fixed according to the recent LHC bounds. The two main mass windows for the SM Higgs mass in the 
$116-145$ GeV and above $466$ GeV can be fixed by natural choices of the coupling constants of the scalar potential, in the range $(-1,1)$.

\begin{widetext}
\appendix
\section{Mass Matrix}
\setcounter{equation}{0}
\setcounter{footnote}{0}
The following five relations correspond to the necessary potential minimum conditions:
\begin{eqnarray}\displaystyle
&&4s^3\lambda_1-2s\mu_1^2 = 0  \nonumber\\
&&2v_L(v_{M_L}-v_{M_R})\mu_5-2v_L\mu_3^2s+2v_Lv_R^2\lambda_5+4v_L^3\lambda_4 = 0 \nonumber\\
&&-2(v_{M_L}-v_{M_R})v_R\mu_5-2v_R\mu_3^2s+2v_L^2v_R\lambda_5+4v_R^3\lambda_4 = 0 \nonumber\\
&&(v_L^2-v_R^2)\mu_5+v_{M_R}\mu_4^2s-2v_{M_L}\mu_2^2+2v_{M_L}v_{M_R}^2\lambda_3+4v_{M_L}^3\lambda_2 = 0  \nonumber\\
&&-(v_L^2-v_R^2)\mu_5+v_{M_L}\mu_4^2s-2v_{M_R}\mu_2^2+2v_{M_L}^2v_{M_R}\lambda_3+4v_{M_R}^3\lambda_2 = 0
\end{eqnarray}

In the basis \{$S_M, \, \phi_L, \, \phi_R,\, M_{N_L}, \, M_{N_R}$\} the squared-mass matrix is given by 
\begin{eqnarray}
M^2_{\alpha, \beta} = \frac{\partial^2 V}{\partial \, \Phi_{\alpha}\partial \, \Phi_{\beta}}.
\end{eqnarray} 

By using the constraints above to express the $\mu_i$ parameters in terms of the v.e.v $v_i$, we arrive at the following matrix:

{\scriptsize
\begin{equation}
\label{11}
M^2=\left(
\begin{matrix}
8s^2\lambda_1 & 0 &  0 & 0 & 0
\cr 
0 & 8v_L^2\lambda_4 & 4v_Lv_R\lambda_5 &  v_L \Delta_1 &  - v_L \Delta_1
\cr 
0 & 4v_Lv_R\lambda_5 &  8v_R^2\lambda_4  & - v_R \Delta_1 &  v_R \Delta_1
\cr
0 & v_L \Delta_1 &  -v_R \Delta_1  &  \Delta_2  
+ 2v_{M_R}^2 (2\lambda_2 - \lambda_3) - 8v_{M_L}^2\lambda_2
 & \Delta_2 + 
2v_{M_L}v_{M_R}(2\lambda_2 + \lambda_3)
\cr 
0 & -v_L \Delta_1 &
v_R \Delta_1 & \Delta_2 +2v_{M_L}v_{M_R}(2\lambda_2 + \lambda_3) &  
\Delta_2 +
2v_{M_L}^2 (2\lambda_2 - \lambda_3) - 8v_{M_R}^2\lambda_2
\end{matrix} \right),
\end{equation}}
\noindent
with the definitions,
$$\Delta_1 =\frac{(v_R^2-v_L^2)(\lambda_5-2\lambda_4)}{(v_{M_R}-v_{M_L})} \qquad {\rm and} \qquad \Delta_2= \frac{(v_R^2-v_L^2)^2(\lambda_5- 2 \lambda_4)}{2(v_{M_R}-v_{M_L})^2}.  $$

\medskip
The previous results can be approximated in the limits $\Delta_1 \rightarrow 0 $ and  $\Delta_2 \rightarrow 0 $. In all our results, no 
fine-tuning conditions are imposed on the scalar potential.

\begin{table}[ht]\label{Dachshund}
\begin{footnotesize}
\centering
\begin{tabular}{|c|c|c|c|c|c|}
\hline
\hline
&    &   &   &  &    \\  
$v_{M_L}$ (TeV) & $Z^{\prime} \rightarrow  \displaystyle\sum\limits_{i=1}^3\, \bar \nu_i \nu_i$   
& $Z^{\prime} \rightarrow \displaystyle\sum\limits_{i=1}^3 \, \bar l_i l_i$
& $Z^{\prime} \rightarrow   \bar N N $ 
& $Z^{\prime} \rightarrow  \displaystyle\sum\limits_{i=1}^6 \,\bar q_i q_i$ 
& $Z^{\prime} \rightarrow  \displaystyle\sum\limits_{i=1}^3 \, (\bar \nu_i N + \bar N \nu_i) $   \\ \hline
&    &   &   &  &    \\  
$1$  & $60\%$   &   $15.9\%$     & $< 10^{-3}\%$   &  $23.2\%$ &  $1.9\%$ \\ 
\hline 
&    &    &  &  &    \\ 
$10$  & $60\%$   &   $16.2\%$     & $< 10^{-6}\%$   &  $23.6\%$ &  $0.02\%$ \\ \hline
\end{tabular}
\end{footnotesize}
\caption{The $Z^{\prime}$ branching-ratios for $M_{Z^\prime}= 1.5$ TeV for  $v_{{M_L}}= 1$ TeV and $10$ TeV. }
\end{table}

\end{widetext}

\medskip
\section*{Acknowledgements}
This work was partially supported by the following Brazilian agencies: CAPES, CNPq and FAPERJ.


\begin{thebibliography}{ABC}
\bibitem{PDG} K.~Nakamura {\it  et al.}, (Particle Data Group), J. Phys. G {\bf 37}, 075021 (2010).
\bibitem{T2K} K.~Abe {\it et al.} (T2K Collaboration), Phys. Rev. Lett. {\bf 107}  041801 (2011). arXiv:hep-ph/1106.2822.
\bibitem{ALT} E.~Ma and D.~Wegman, Phys. Rev. Lett. {\bf 107}, 061803 (2011). arXiv:hep-ph/1106.4269; G.~Altarelli and F.~Feruglio, 
Rev. Mod. Phys. {\bf 82}, 2701 (2010). arXiv:hep-ph/1002.0211; N.~Haba, T.~Horita, K.~Kaneta and Y.~Mimura, arXiv:hep-ph/1110.2252v1.
\bibitem{MIN} A.~A.~Aguilar-Arevalo {\it et al.} (MiniBooNe Collaboration), Phys. Rev. Lett. {\bf 105}, 181801 (2010), 
arXiv:hep-ex/1007.1150.
\bibitem{GIU} C.~Giunti and M.~Laveder, arXiv:hep-ph/1107.1452; J.~Kopp, M.~Maltoni and T.~Schwetz, arXiv:hep-ph/1103.4570.
\bibitem{JAP} N.~Haba and K.~Tsumura, JHEP 1106, 068 (2011). arXiv:hep-ph/1105.1409.
\bibitem{JCP} J.~C.~Pati and A.~Salam, Phys. Rev. D {\bf 10}, 275 (1974), Erratum-ibid. D {\bf 11} 703 (1975);
 R.~N.~Mohapatra and J.~C.~Pati, Phys. Rev. D {\bf 11}, 566 (1975); G.~Senjanovi\u c and R.~N.~Mohapatra, Phys. Rev. D {\bf 12},
 1502 (1975); R.~N.~Mohapatra and R.~E.~Marshak, Phys. Lett. B {\bf 91}, 222 (1980). An extensive list of earlier references can be 
found in  R.~N.~Mohapatra and P.~B.~Pal, "Massive Neutrinos in Physics and Astrophysics", World Scientific, Singapore, 1998. 
\bibitem{CMP} D.~Chang, R.~N.~Mohapatra, M.~K.~Parida, Phys. Rev. D {\bf 30} 1052 (1984).
\bibitem{MMV} R.~N.~Mohapatra, Phys. Rev. Lett. {\bf 56}, 561 (1986); R.~N.~Mohapatra and J.~W.~F.~Valle, Phys. Rev. D {\bf 34},
 1642 (1986).
\bibitem{MART} H.~Martinez, A.~Melfo, F.~Nesti and G.~Senjanovi\u c, Phys. Rev. Lett. {\bf 106} (2011) 191802.
arXiv:hep-ph/1101.3796.
\bibitem{FERM} The TEVNPH Working Group for the CDF and D0 Collaborations, arXiv:hep-ex/1109.3357v1.
\bibitem{ATLAS} ATLAS Collaboration, arXiv:hep-ex/1109.3357v1.
\bibitem{CMS} CMS Collaboration, CMS PAS HIG-11-015.
\bibitem{DEALM} F.~M.~L.~de~ Almeida, Y.~A.~Coutinho, J.~A.~Martins Sim\~oes, A.~J.~Ramalho, L.~Ribeiro Pinto, 
S.~Wulck and M.~A.~B.~do Vale, Phys. Rev. D {\bf 81} 053005 (2010). arXiv:hep-ph/1001.2162.
\bibitem{RODE} W.~Rodejohann, Int. J. Mod.\ Phys. E {\bf 20}, 1833 (2011) arXiv:hep-ph/1106.1334.
\bibitem{CAR} L.~M.~Carpenter, M.~ Dine and G.~Festuccia, Phys. Rev. D {\bf 80} (2009) 125017.
\bibitem{ATL} The Atlas Collaboration, arXiv:hep-ex/1108.1582.
\bibitem{COMP} E.~Boos {\it et al.}, (CompHEP Collaboration), CompHEP 4.4: Automatic computations from Lagrangians to events,
Nucl. Instrum. Meth. A {\bf 534}, 250 (2004).
\bibitem{CHART} S.~Chatrchyan {\it et al.}, CMS Collaboration, JHEP {\bf 1106}, 077 (2011). arXiv:hep-ex/1104.3168.
\bibitem{ALM} F.~M.~L.~Almeida, Y.~A.~Coutinho, J.~A.~Martins Sim\~oes and M.~A.~B.~do Vale, Phys. Rev. D {\bf 62}, 075004 
(2000); O.~Panella, M.~Cannoni, C.~Carimalo and Y.~N.~Srivastava, Phys. Rev. D {\bf 65}, 035005 (2002).
\bibitem {PON} J.~A.~Martins Sim\~oes, J.~Ponciano, Eur. Phys. J. direct C {\bf 30}, 007 (2003), Eur. Phys. J. C 32S1, 91 (2004). 
\end{thebibliography}
\end{document}